# Changing Electricity Markets: Quantifying the Price Effects of Greening the Energy Matrix

Emanuel Kohlscheen and Richhild Moessner [1,2]

## Abstract

We analyse the drivers of European Power Exchange (EPEX) wholesale electricity prices between 2012 and early 2022 using machine learning. The agnostic random forest approach that we use is able to reduce in-sample root mean square errors (RMSEs) by around 50% when compared to a standard linear least square model – indicating that non-linearities and interaction effects are key in wholesale electricity markets. Out-of- sample prediction errors using machine learning are (slightly) lower than even in-sample least square errors using a least square model. The effects of efforts to limit power consumption and green the energy matrix on wholesale electricity prices are first order. CO2 permit prices strongly impact electricity prices, as do the prices of source energy commodities. And carbon permit prices' impact has clearly increased post-2021 (particularly for baseload prices). Among energy sources, natural gas has the largest effect on electricity prices. Importantly, the role of wind energy feed-in has slowly risen over time, and its impact is now roughly on par with that of coal.

[1] Both authors are Senior Economists from the Bank for International Settlements, Centralbahnplatz 2, 4002 Basel, Switzerland. *E-Mail addresses*: emanuel.kohlscheen@bis.org, richhild.moessner@bis.org.
[2] We are grateful to Adam Cap and Emese Kuruc for help in getting the data. The views expressed in this paper are those of the authors and do not necessarily reflect those of the Bank for International Settlements.

## 1. Introduction

Price incentives are central to foster the transition to a less carbon intensive and more sustainable economy. Both higher carbon taxes and tradable emission permits lead to greater internalization of the adverse environmental externalities which are generated by economic activities. Ideally, proper pricing of carbon emissions would dampen overall energy demand. At the same time, it would tilt demand from more dirty energy sources towards cleaner and renewable ones over time.

This study presents a computationally intensive machine learning based model that is able to explain one-day ahead wholesale electricity prices in the European Power Exchange (EPEX) over the past 10 years relatively well. It does so based on a comprehensive list of conditioning factors, which include prices of major energy commodities, weather conditions (i.e. temperature and sun hours), financial factors, inflation, wind power feed-ins and the price of carbon permits.

The agnostic benchmark random forest model that is presented is based on an ensemble of 1,000 regression trees. The model reduces in-sample root mean square errors (RMSEs) by around 50% when compared to a linear least square model. This fact highlights that non-linearities and interaction effects are key in wholesale electricity markets – an observation that is confirmed by the occasional jumps and spikes that often characterize these markets which require continuous balancing of supply and demand of a non-storable resource (see Bierbrauer et al (2007)).[3] Out-of-sample prediction errors are also (slightly) lower than even in-sample least square errors. Overall, our analysis reveals a strong effect of price momentum, reflecting a general tendency of price variations to reverse.

Importantly, the effects of efforts to limit power consumption and green the energy matrix on prices are found to be material. CO2 permit prices strongly impact EPEX wholesale electricity prices, as do the prices of key source energy commodities such as oil, coal and natural gas. Further, we confirm that the impact of carbon permit prices has clearly increased post-2021 (particularly for baseload prices). Among energy sources, we find that natural gas has the largest impact on wholesale electricity prices. Importantly, the role of wind energy feed-in has slowly risen over time. Its impact is already roughly on par with that of coal.

**Relation to the literature.** This paper adds to the literature by focusing on the aspects that are related to the challenges of greening the energy matrix. It does so by using a widely established machine learning technique to model electricity prices.[4] This is important, particularly in view of the fact that random forests (Breiman al (1984) and Breiman (2001)) are a disciplined and flexible non-parametric method that can accommodate the highly non-linear dynamics and interactions that often occur in electricity markets.[5] [6] In terms of technique, the energy market paper that is closest to ours is the recent paper by Castañeda et al (2021) who also used the random forest approach –in their case to forecast natural gas prices. One important innovation of our study is that, beyond analyzing drivers of electricity prices, we elicit partial effects of selected variables, thus enhancing economic interpretation and quantification of the effects.

---

[3] By and large, electricity storage is restricted to pumped hydro. For a systematic study of the effects of other storage innovations in this field see Lazkano et al (2017).
[4] For a discussion on the use of machine learning in economic applications see Kleinberg et al (2015) and Mullainathan and Spiess (2017).

Of course, our study benefits from and builds on several articles on electricity prices. Here we highlight a few of them. Bierbrauer et al (2007) and Frömmel et al (2014) are some of the earliest analyses of the EPEX electricity market.[7] Both studies were based on econometric models, as is most of the literature on the topic that followed. The former found that regime switching models perform relatively well in explaining price patterns. The later focused on forecasting prices using GARCH models.[8] In terms of economic results, the strong price reversal effect that we find is in line with the mean-reversion property that had been documented earlier. Bierbrauer et al (2007) related this to the switching on (off) of energy sources with high marginal costs when demand increases (decreases).

Ketterer (2014) was one of the first to investigate the effects of renewable energy on wholesale electricity prices. She focused on the effects of wind power and found that its supply impacts prices and that – due to their intermittent nature – supply of renewables might increase price volatility. Her study was also based on a GARCH model, and at the time did not include controls for other energy sources.

To the best of our knowledge, our inclusion of carbon permit prices is novel. This inclusion is particularly relevant in view of their very sharp recent movements and the fact that they represent a key element in the transition strategy to a greener economy.[9] Our model is based on the most comprehensive list of economic control variables so far – given that we are able to include daily data on the bulk of energy sources, carbon permits, temperatures, sun hours, as well as financial factors.

---

[5] For instance, Mosquera-López and Nursimulu (2017) uncover non-linearities, and analyze prices using a threshold based econometric model.
[6] Fernandez-Delgado et al (2014) compared the performance of 179 classifier models across 121 datasets, and found (the relatively simple) random forests to be the top performers.
[7] Addressing a different question, Kath and Ziel (2018) find substantial profit opportunities from intraday trading in electricity spot markets.
[8] Longstaff and Wang (2004) studied spot and one-day ahead electricity prices in Pennsylvania, New Jersey and Maryland between 2000 and 2002. Interestingly, they find that one-day forward prices are less volatile than expected spot prices.
[9] On this aspect, see Kohlscheen et al (2021). They find that higher carbon taxes and prices of permits significantly reduce $CO_2$ emissions across countries and time.

## 2. Background and Data

To analyze the drivers of the electricity market, we focus on the one-day ahead EPEX wholesale electricity price.[10] These prices are quoted for every 15-minute interval. For the purposes of this study, we average prices for each working day. Figure 1 shows the evolution of monthly averages for these since the beginning of our sample, i.e. from Jan 2nd, 2012. Monthly averages are taken as 22-day rolling averages. The baseload price over the 10-year period averages 43.12 €/MWh, with a standard deviation of 20.55, while the peakload price averages 47.65 €/MWh, with a standard deviation of 24.28. As Figure 1 shows, prices spiked from the second half of 2021, as the price of source commodities rose strongly. In December 2021, the baseload price was at 228 €/MWh and the peakload price at 269 €/MWh.[11]

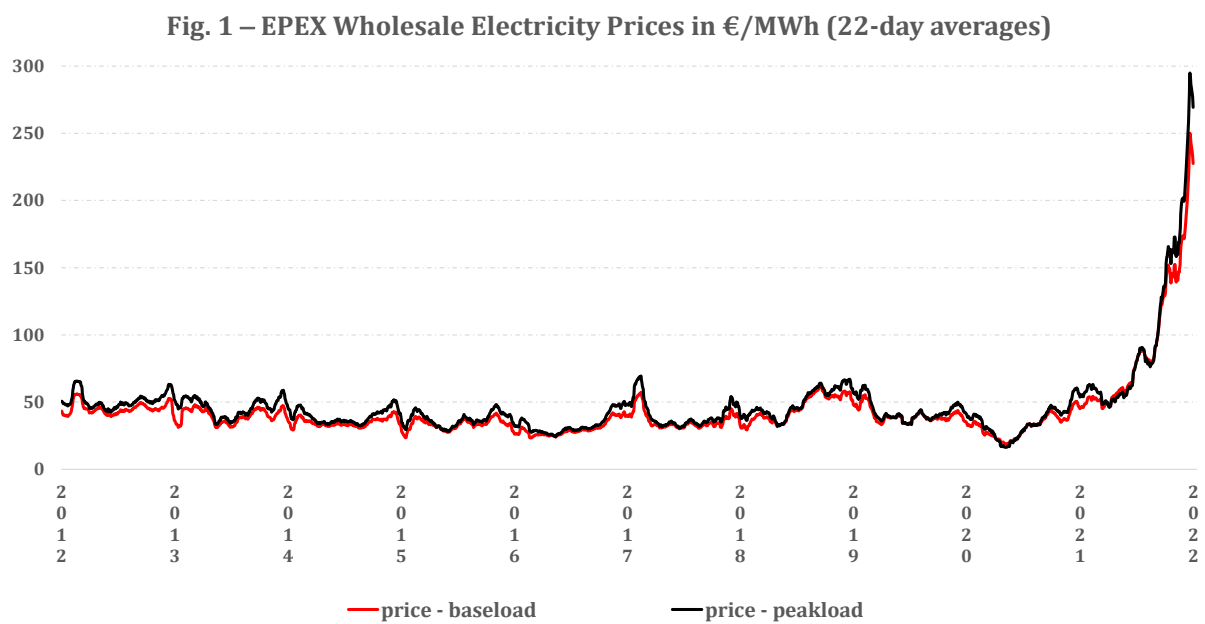

Fig. 1 – EPEX Wholesale Electricity Prices in €/MWh (22-day averages)

Besides energy source commodities (such as natural gas, oil and coal), energy from renewable sources is also fed into the system. Wind power is the most important of these, accounting for 23.3% of energy production in 2021. Solar energy is the second most important source (8.6%), followed by biomass (7.9%).[12] Appendix Figure 1 shows the evolution of the wind power feed-in in gigawatts. The contribution of wind is highly volatile, depending on weather conditions (see Ketterer (2014)). That said, it has the useful property that supply is generally higher during the winter months, when energy demand for heating is highest.

---

[10] EPEX operates in Austria, Belgium, Denmark, Finland, France, Germany, Great Britain, Luxembourg, the Netherlands, Norway, Poland, Sweden and Switzerland.

[11] Duso and Szűcz (2017) estimate that between 50 and 70% of producer cost changes are eventually transmitted to final consumers.

[12] See Statistisches Bundesamt. Coal accounted for 23.7% and natural gas for 16.8% of production.

Global warming could reduce wind speeds in northern Europe by reducing the temperature difference between the equator and North pole which is driving these winds. Wind speeds over western, central and northern Europe are predicted by the Intergovernmental Panel on Climate Change to drop by up to 10% in the summer months by 2100 based on 1.5°C warming (Bernard (2021). Such an effect from global warming would also make the contribution from wind energy more volatile, and could already have been a factor in Europe in the summer of 2021, contributing to higher electricity prices then.

A further noteworthy development in energy markets over the time period under consideration has been the rise in CO2 emission permits prices, as shown in Figure 2. These permits are traded in the Emissions Trading System (ETS), which was launched in the European Union in 2005. Besides providing incentives for reductions in overall energy demand, higher permit prices have become a mechanism to steer energy sourcing towards more environmentally friendly options. Parry (2020) provides a detailed discussion of carbon pricing options in the context of the European Union. His analysis concludes that higher emissions prices in the ETS would create the largest welfare gain.[13]

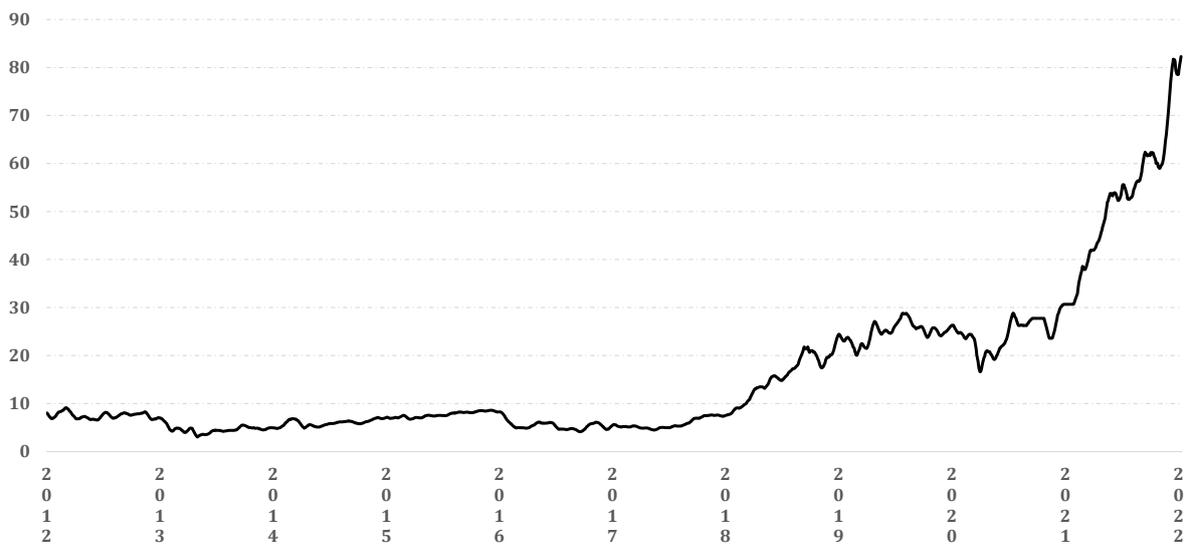

Fig. 2 – Permit price in € per metric tonne (22-day average)

## 3. Methodology and Model Validation

### 3.1 Data

The price evolutions reported in the previous sections already suggest that non-linearities, and possibly interactions between different driving factors, are likely to be important in EPEX electricity prices. In particular, non-linearities may kick in when supply of a given energy source comes close to practical supply limits, causing prices

---

[13] For a more general discussion, see Stiglitz (2019) and van der Ploeg and Rezai (2019).

to spike. For this reason, we chose to model the electricity market using machine learning.

More specifically, we rely on random forests (see Breiman al (1984) and Breiman (2001)). These are an ensemble of regression trees. This widely used non-parametric tool is highly flexible and has the main advantage that it provides an agnostic, but still disciplined way to disentangle possibly non-linear and/or complex relations between driving factors and the outcome variable (in our case the wholesale electricity price).

Regression trees treat outcome prediction as a classification problem. The computational algorithm searches for the predictor variable and an associated threshold value on which to split the dataset of observations. The optimal split is the one that minimizes the sum of squared residuals of errors after the split. The resulting nodes and subsamples are then split again, successively, until a stopping criterion is reached (i.e. the maximum depth of the regression tree or, put differently, the minimum number of observations per splitting node). Eventually, the algorithm predicts outcome values for out-of-sample data points based on the outcome in the regression tree in which that observation would fall.

In our case, we predict the 22-day change in the EPEX electricity price, based on the information of 12 predictor variables (features). These are the one-month lagged change, to capture autocorrelation or mean-reversion patterns, and the 22-day changes in the international prices of Brent oil, Rotterdam coal and ICE natural gas prices, all converted to euros. To these we add the changes in CO2 permit price, the daily wind power feed-in, and average sun hours in Berlin (Tempelhof) weather station. The latter is meant to be a crude proxy of total photovoltaic energy that is fed into the grid. Throughout, we control for the average daily temperature, and the day of the week (to capture eventual seasonalities in price effects). Last, we control for the prevailing interest rate, for the VIX market volatility index traded on the Chicago Board Options Exchange (CBOE) and for the consumer price inflation (CPI) index. The complete data description is provided in the Appendix.

Note that our data is daily, but since we are primarily interested in medium-term fluctuations, the analysis that we present is based on 22-day variations or averages (i.e. monthly). We use this data-driven method to explain current movements in electricity prices at each point in time.[14]

### 3.2 Random Forests

Random forests rely on averages over a large number of regression trees. This has the advantage that it greatly increases the robustness of the model by reducing the variability of predictions and raising out-of-sample performance.[15] As can be seen in

---

[14] The adaptation of the model for longer-horizon forecasting applications is straightforward, and follows directly from the change of the outcome variable (in the spirit of the local projection method of Jordà (2005)).

[15] See Breiman (2001) and Friedman et al (2009). Mentch and Zhou (2020) find that "*the additional randomness injected into individual trees serves as a form of implicit regularization, making random forests an ideal model in low signal to noise (SNR) settings.*"

Figure 3, mean square errors of the model reduce rapidly as the number of trees grows (horizontal axis increases).

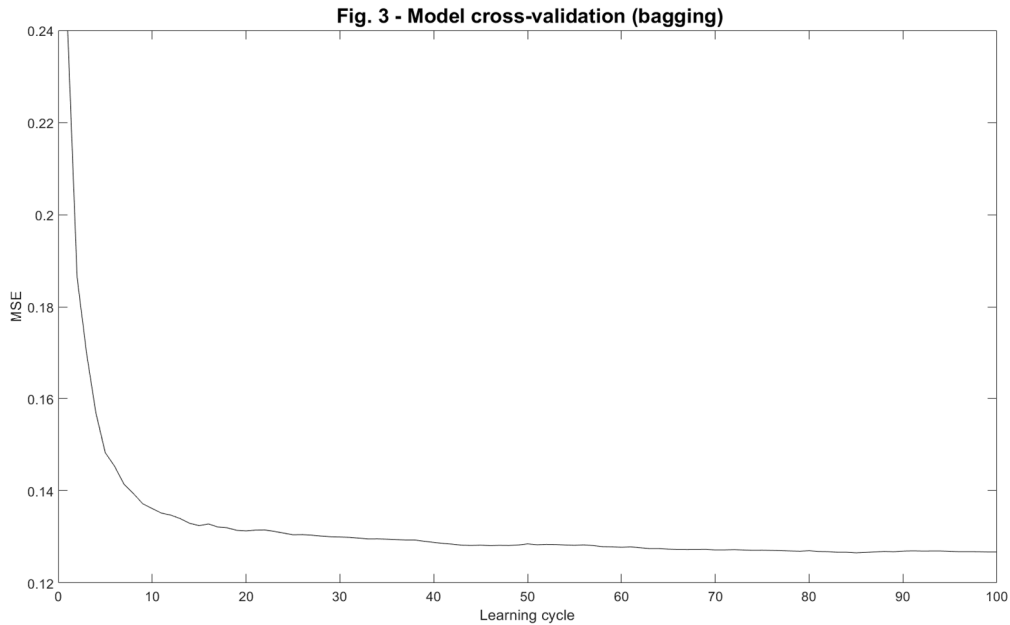

Table 1 shows root mean square errors (RMSEs) of the resulting models for alternative stopping criteria (in the first column). Throughout, roughly 1/3 of the observations are randomly excluded from the sample, and a tree is grown based on the remaining 2/3 of observations. This is then repeated many times (in our case either 100 or 1,000), to produce the random forest.

What the table shows is that deeper trees (i.e. lower min obs/splitting node), lead to better model fits. The caveat is that this may lead to overfitting of the model. Yet, the cost of overfitting is ultimately an empirical issue, and depends on the application. In our case, we find that the cost of overfitting is very small: the out of sample ("out-of-bag") performance barely worsens by the selection of a deeper tree. For this reason, we select 10 as the minimum number of observations per splitting node, with 1,000 trees as our benchmark model.

**Table 1 – RMSEs for Base- and Peakload Electricity Prices vs. Minimum Size of Parent Splitting Nodes**

| | | root mean square errors (RMSEs) | | | | RMSE ratio (rel. OLS) |
|---|---|---|---|---|---|---|
| base | | base | | | | |
| | | econometrics (in sample) | | ML | | |
| min obs/splitting node | no. of reg trees | AR(1) | OLS | in sample | out of bag | |
| 5 | 100 | 0.404 | 0.339 | 0.144 | 0.316 | 0.425 |
| | 1,000 | .. | .. | 0.142 | 0.311 | 0.418 |
| 10 | 100 | .. | .. | 0.169 | 0.314 | 0.498 |
| | 1,000 | .. | .. | 0.168 | 0.312 | 0.495 |
| 20 | 100 | .. | .. | 0.206 | 0.318 | 0.607 |
| | 1,000 | .. | .. | 0.203 | 0.314 | 0.599 |
| 30 | 100 | .. | .. | 0.227 | 0.319 | 0.669 |
| | 1,000 | .. | .. | 0.225 | 0.317 | 0.664 |
| 40 | 100 | .. | .. | 0.240 | 0.320 | 0.709 |
| | 1,000 | .. | .. | 0.240 | 0.319 | 0.708 |
| peak | | peak | | | | |
| | | econometrics (in sample) | | ML | | |
| min obs/splitting node | no. of reg trees | AR(1) | OLS | in sample | out of bag | |
| 5 | 100 | 0.404 | 0.380 | 0.157 | 0.358 | 0.414 |
| | 1,000 | .. | .. | 0.155 | 0.352 | 0.407 |
| 10 | 100 | .. | .. | 0.188 | 0.360 | 0.495 |
| | 1,000 | .. | .. | 0.185 | 0.353 | 0.487 |
| 20 | 100 | .. | .. | 0.227 | 0.357 | 0.597 |
| | 1,000 | .. | .. | 0.226 | 0.355 | 0.594 |
| 30 | 100 | .. | .. | 0.254 | 0.360 | 0.667 |
| | 1,000 | .. | .. | 0.251 | 0.357 | 0.660 |
| 40 | 100 | .. | .. | 0.268 | 0.360 | 0.705 |
| | 1,000 | .. | .. | 0.268 | 0.359 | 0.706 |

Note: Sample period is from Jan 2012 to Jan 2022.

The benchmark random forest model (with 1,000 trees) delivers a 50.5% reduction in RMSE relative to the OLS model in-sample for the baseload price, and a 51.3% reduction for the peakload price.[16] In other words, the machine learning model provides a more accurate description of how explanatory variables (or features) relate to electricity prices. What is more, the out of sample (out-of-bag) RMSEs of 0.312 (baseload) and 0.353 (peakload) are slightly below even the respective (in sample) RMSEs of a naïve AR(1) model and of the OLS model which contains the same 12 explanatory variables.[17] All in all, this confirms the relative high performance of the random forest model.

---

[16] I.e. (1 – 0.495) and (1 – 0.487).
[17] For baseload prices, out-of-bag RMSEs are 22.8% below (in-sample) AR(1) RMSEs and 8% below OLS RMSEs. For peakload prices, 22.6% and 7.1%, respectively.

## 4. Results for the Full Sample

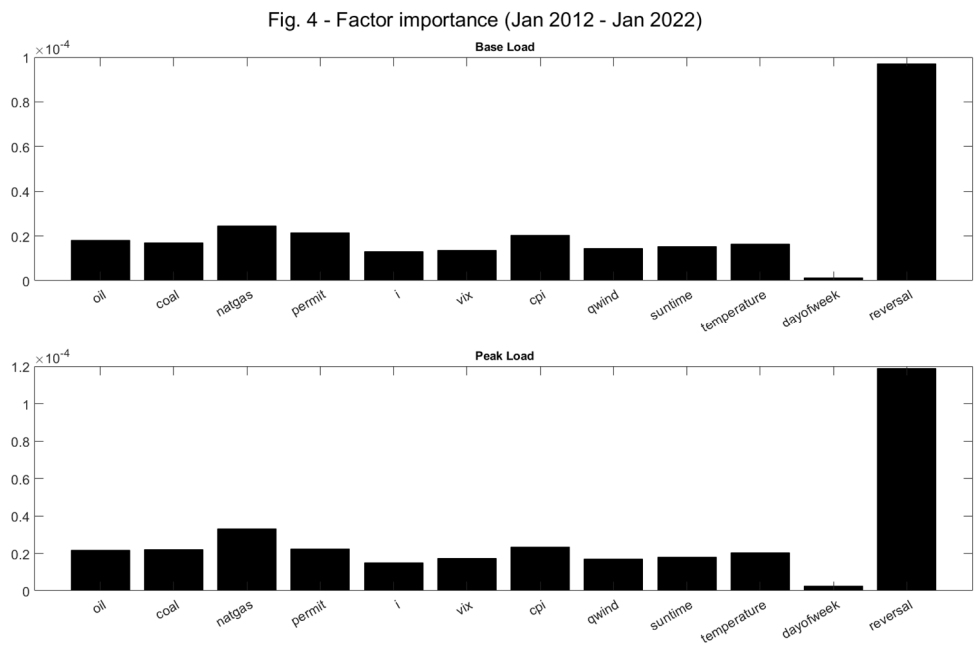

Fig. 4 - Factor importance (Jan 2012 - Jan 2022)

The relative contribution of each explanatory factor to the reduction of mean square errors is shown by the predictor importance statistics in Figure 4. The most relevant factor in explaining 22-day price variations is the mean reversal factor (denoted by reversal, both for base- and for peakload). This is just the 22-day lag of the wholesale price changes. In the equivalent OLS model, this lagged dependent variable has a coefficient of −0.5, indicating that – all else equal – about half of price increases in one month are typically reversed in the following month.

The normalized factor importances reveal that natural gas prices are the most important drivers of EPEX wholesale prices. This is likely related to the fact that natural gas is commonly the marginal energy source.

**Table 2 - Normalized factor importance**

|  | permit | oil | coal | nat gas | i | vix | cpi | qwind | suntime | temp | day_week | reversal |
|---|---|---|---|---|---|---|---|---|---|---|---|---|
| Base | 0.079 | 0.066 | 0.062 | 0.090 | 0.048 | 0.050 | 0.074 | 0.053 | 0.056 | 0.060 | 0.005 | 0.357 |
| Peak | 0.068 | 0.065 | 0.066 | 0.100 | 0.045 | 0.052 | 0.071 | 0.051 | 0.054 | 0.062 | 0.008 | 0.358 |

Note: Normalized so that sum of all predictor importances equals 1.

Figure 5 depicts the predicted interrelation of the two main drivers of electricity prices: carbon permits and natural gas. Each predicted monthly electricity price variation (vertical axis) is based on the average of the 1,000 regression trees. These partial effects are computed holding all other explanatory factors fixed at their sample means.[18] Electricity prices increase when both CO2 permit and natural gas prices increase, *ceteris paribus*.

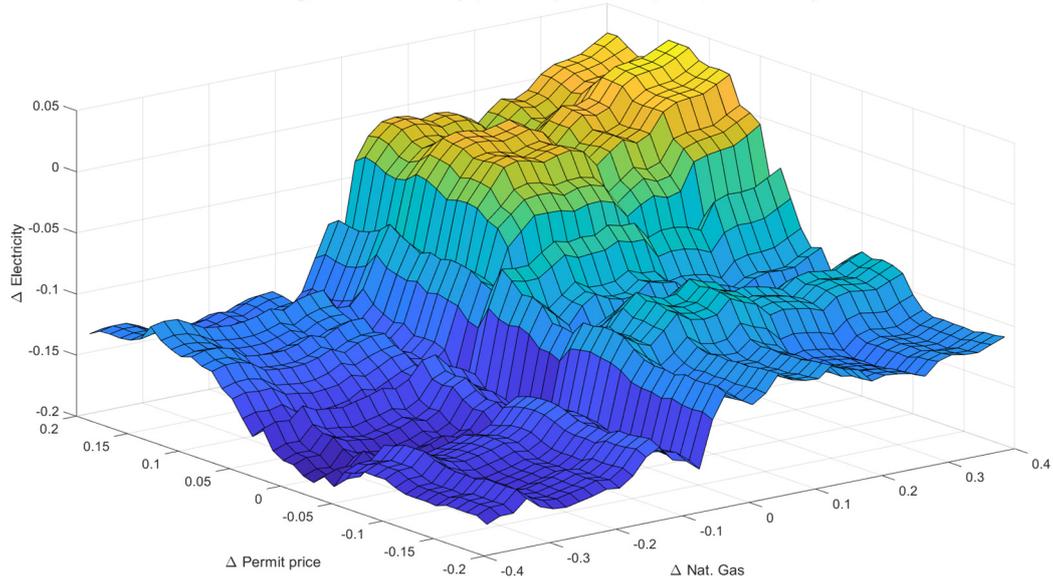

## 5. Recent Sample

Figure 6 shows the contribution of each factor when the sample is restricted to start only in January 2021. Normalized factor importances are depicted in Table 3. What stands out in this more recent sample is that carbon permits have clearly increased their imprint on wholesale electricity prices.

---

[18] In other words, this is roughly equivalent to coefficients in a linear regression (partial effect).

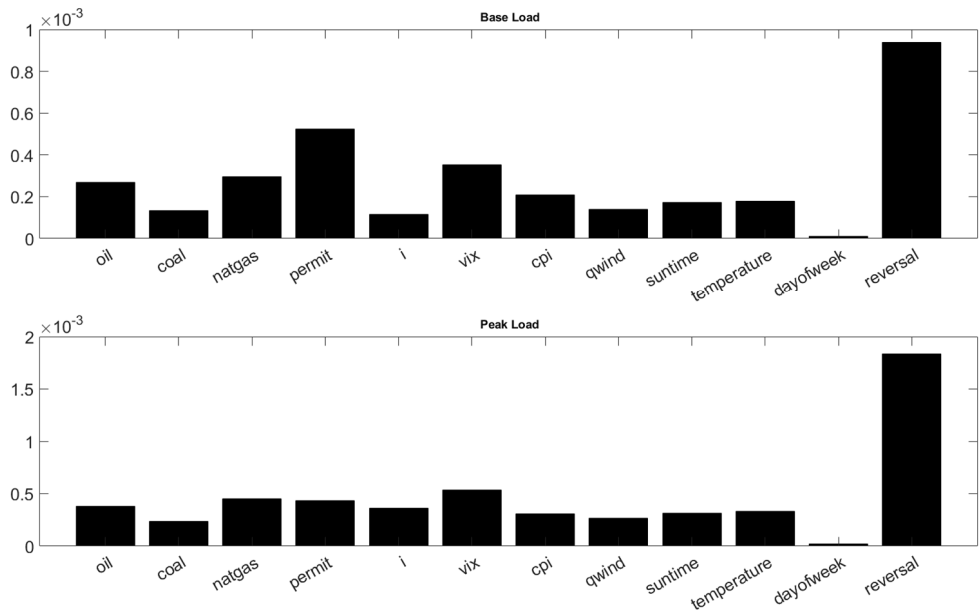

Fig. 6 - Factor importance since Jan 2021

Last but not least, the relevance of renewables has slowly increased over time – as the normalized wind factor has drifted up. During the last year, its contribution has been roughly on par with that of coal. The positive factor for sun time, while correlated with more general weather factors, might also be an indication that the relevance of solar energy has increased at the margin.

**Table 3 – Normalized factor importance (subsamples)**

|  | permit | oil | coal | nat gas | i | vix | cpi | qwind | suntime | temp | day_week | reversal |
|---|---|---|---|---|---|---|---|---|---|---|---|---|
| 2012-2022 | | | | | | | | | | | | |
| Base | 0.079 | 0.066 | 0.062 | 0.090 | 0.048 | 0.050 | 0.074 | 0.053 | 0.056 | 0.060 | 0.005 | 0.357 |
| Peak | 0.068 | 0.065 | 0.066 | 0.100 | 0.045 | 0.052 | 0.071 | 0.051 | 0.054 | 0.062 | 0.008 | 0.358 |
| since 2017 | | | | | | | | | | | | |
| Base | 0.089 | 0.064 | 0.061 | 0.082 | 0.043 | 0.051 | 0.083 | 0.055 | 0.053 | 0.070 | 0.004 | 0.344 |
| Peak | 0.073 | 0.073 | 0.068 | 0.095 | 0.043 | 0.053 | 0.069 | 0.054 | 0.058 | 0.067 | 0.005 | 0.343 |
| since 2021 | | | | | | | | | | | | |
| Base | 0.132 | 0.079 | 0.061 | 0.087 | 0.040 | 0.068 | 0.063 | 0.056 | 0.054 | 0.062 | 0.003 | 0.295 |
| Peak | 0.080 | 0.084 | 0.059 | 0.085 | 0.054 | 0.063 | 0.054 | 0.057 | 0.058 | 0.061 | 0.004 | 0.340 |

Note: Normalized so that sum of all predictor importances equals 1.

## 6. Conclusion

Summing up, this article presents a model of wholesale electricity prices in the European Power Exchange. The model is based on the most comprehensive list of conditioning factors so far. The model can easily be adapted for longer-term forecast exercises. Here we focused on averages for baseload and peakload prices, leaving analysis for intraday patterns for future research efforts.

Our results show that green policies have systematic first-order effects on electricity prices. Wholesale electricity prices show strong (negative) serial dependence, reflecting a general tendency to revert to the mean. Among commodities, natural gas prices have the strongest effect on electricity prices overall, followed by oil and coal – possibly due to the role of natural gas as the marginal supply.

$CO_2$ emission permits strongly impact electricity prices, and their effect has surged recently. Wind-power feed-ins as well as sun time also impact prices, highlighting the growing relevance of renewable energy sources.

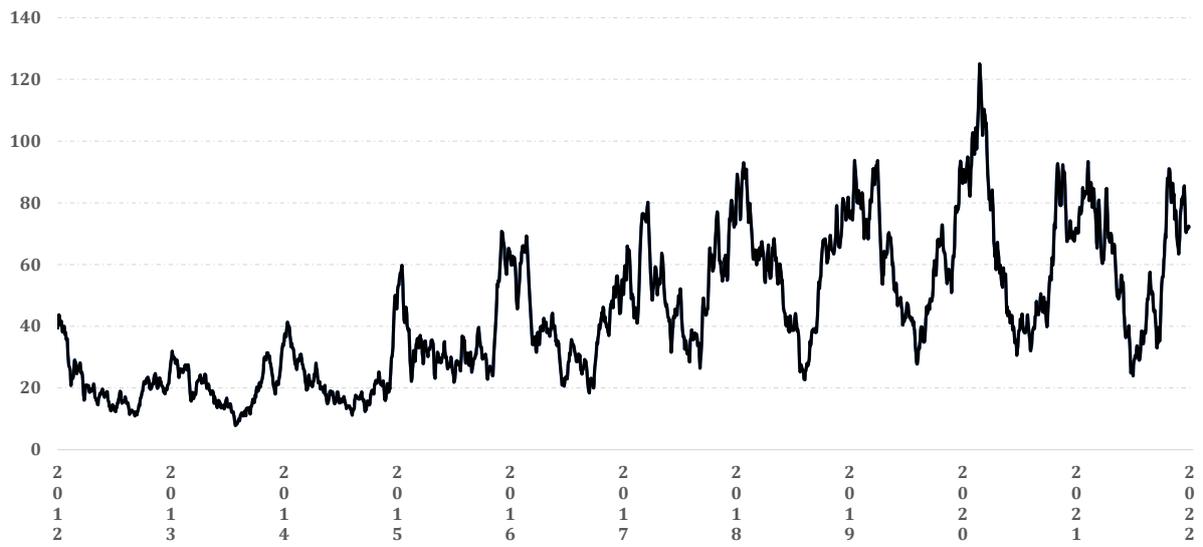

Fig. A1 – Wind Power Feed-In in GW (22-day average)

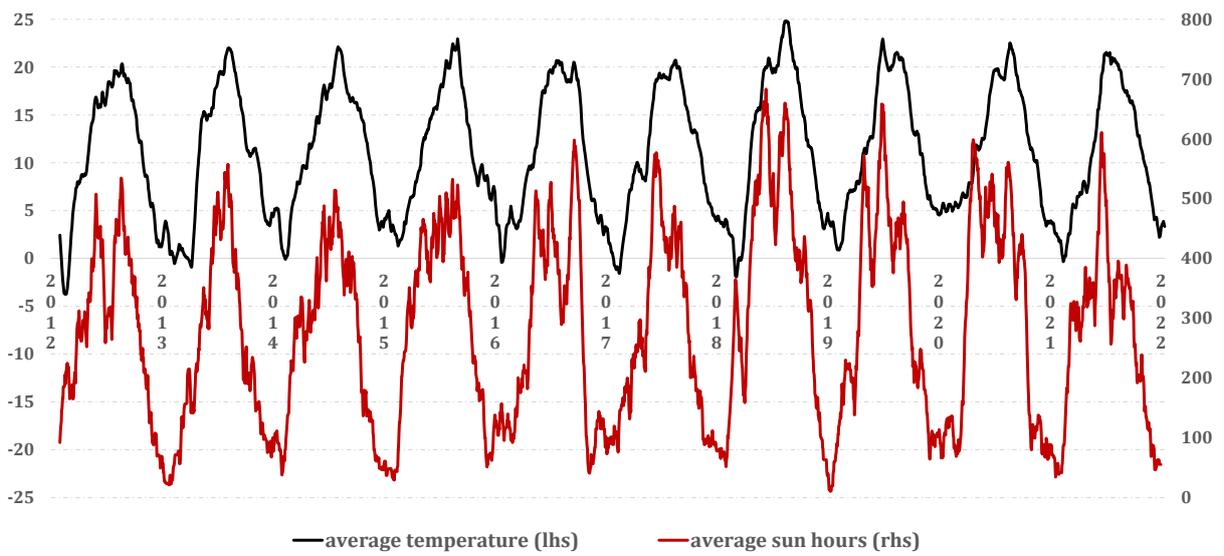

Fig. A2 – Temperature and Sun Hours

## Data Appendix

The database on which this study is based comprises 2,624 daily observations of the following variables:

- EPEX spot day-ahead baseload and peakload electricity prices, from Bloomberg (LPXBHRBS Index and LPXBHRPK Index). Prices in € per MWh.
- permit: $CO_2$ emission permit prices, from Bloomberg (EECXSYR1 Index). Prices in € per metric tonne.
- oil: oil price per barrel, Brent type, from Bloomberg (CO1 Comdty).
- coal: Coal, Rotterdam price, from Bloomberg (XA1 Comdty).
- natgas: ICE UK natural gas price, from Bloomberg (FSFUM1 Index). US$ per mbtu.
- qwind: Wind power feed-in, converted to daily average, from TenneT transmission system operator. In GW.
- temperature and suntime: Average daily temperature and sunlight hours in Berlin, Tempelhof, from Meteostat.
- EUR/USD rate, from Bloomberg. For conversion of dollar prices to euros.
- i: EONIA interest rate for euro area, from Bloomberg.
- vix: VIX volatility index traded on the CBOE, from Bloomberg.
- cpi: Consumer price index (CPI), Germany, from the OECD. Interpolated to daily frequency.

The time period of the sample is 2 January 2012 – 11 January 2022.